# Economical and versatile subunit design principles for self-assembled DNA origami structures

*Wei-Shao Wei[1,2], Thomas E. Videbæk[1,2], Daichi Hayakawa[1,2], Rupam Saha[1,2], W. Benjamin Rogers[1,2], Seth Fraden\*[1,2]*

[1] Martin A. Fisher School of Physics, Brandeis University, Waltham, MA 02453, USA

[2] Materials Research Science and Engineering Center (MRSEC), Brandeis University, Waltham, MA 02453, USA

Self-assembly of nanoscale synthetic subunits is a promising bottom-up strategy for fabrication of functional materials. Here, we introduce a design principle for DNA origami nanoparticles of 50-nm size, exploiting *modularity*, to make a family of versatile subunits that can target an abundant variety of self-assembled structures. The subunits are based on a 'core module' that remains constant among all the subunits. Variable 'bond modules' and 'angle modules' are added to the exterior of the core to control interaction specificity, strength and structural geometry. A series of subunits with designed bond/angle modules are demonstrated to self-assemble into a rich variety of structures with different Gaussian curvatures, exemplified by sheets, spherical shells, and tubes. The design features *flexible joints* implemented using single-stranded angle modules between adjacent subunits whose mechanical properties, such as bending elastic moduli, are inferred from cryo-EM. Our findings suggest that incorporating a judicious amount of flexibility in the bond provides error tolerances in design and fabrication while still guaranteeing target fidelity. Lastly, while increasing flexibility could introduce greater variability and potential errors in assembly, these effects can be counterbalanced by increasing the number of distinct bonds, thereby allowing for precise targeting of specific structural binding angles within a broad range of configurations.

**INTRODUCTION**

Self-assembly of elementary nanoscale subunits into complex supramolecular structures is a hallmark of living systems,[1–7] and exploiting the design principles of living systems to create new synthetic materials has been a longtime goal of bioinspired material science.[8–12] In this pursuit, the soft-matter field has identified a set of desirable attributes for self-assembling subunits, colloquially referred to as 'patchy particles'.[13–18] A patchy particle allows the user to design the particle's form and to specify how the particle interacts with other particles. Specifically, the ideal particle allows one to independently optimize the shape and size of the particle, the valency or number of bonds, bond angle, bond specificity and bond strength [Figure 1A]. The need to create customized patchy particles for assembling desired structures remains a key challenge, given the significant effort required to synthesize a single particle that possesses all these attributes.

We recently accomplished this goal by creating patchy particles made from DNA origami with all the attributes listed above and used them to assemble capsid shells[19,20] and tubules.[21] The bond angle was controlled by beveling the edges of the subunits, which specified the local curvature of the resultant self-closing assemblies [Figure 1B]. The bond strength and selectivity were achieved by fabricating complementary lock-and-key domains on the sides of the triangular blocks. However, this approach requires costly re-design each time one makes new subunits with a different bond angle or different interaction specificity.

Here, we introduce a simple and economical design principle, *modularity*, wherein an individual subunit is subdivided into parts – a *universal* 'core module' and two sets of *variable* 'angle modules' and 'bond modules' that encode the local curvature and interaction information, respectively [Figure 1C]. Experimentally, we realize the idea using DNA origami.[22,23] The triangular core module has a maximum valency of three, with bonds forming along each triangular face [Figure 1D]. On each of the three core faces, eight single-stranded DNA (ssDNA) oligos are extended (extruding from the core surface) that contain the bond and angle modules. The portion closer to the core, denoted as the angle module, consists of different lengths of polythymidine (poly-T) nucleotides, which are chosen to control the local curvature of the target assembly. The portion farther from the core, denoted as the bond module, enables programming of the binding specificity and the binding strength between subunits via the base sequences.

Crucially, the designs feature *flexible joints* between pairs of subunits, in large part because of using ssDNA in the angle module. Using cryogenic electron microscopy (cryo-EM) and multi-body refinement, we quantify the bond-angle distribution in subunit pairs. This technique was initially developed to reconstruct molecular motion of proteins with multiple flexible complexes.[24,25] Counterintuitively, we hypothesize that flexible joints offer two advantages. First, flexible joints are expected to tolerate errors in design and fabrication of subunits. Second, large flexibility could make it easier for the two subunits to come into a configuration that results in a bond, increasing the on-rate. In the cases when the joint flexibility is large enough to allow undesirable formation of off-target structures, selectivity of the target structure can then be re-established by increasing the number of distinct bond modules.[26–29]

Using this modular design with flexible joints, we demonstrate the generation of a rich set of distinct subunits. Once the core design is established, each subunit variant requires at most 12% oligo modifications compared to redesigning from scratch. This design principle not only significantly reduces the design and synthetic effort for variants but also makes the platform readily accessible to researchers without expertise in DNA origami design. The core, which contains all the origami-designed portions, remains invariant, while the variable angle and bond modules are simple to design. Here we showcase manufacturing of subunits that self-assemble into various structures, exemplified by ones with zero-Gaussian curvature, *e.g.* 2D tiles and short cylindrical tubes, and positive-Gaussian curvature, *e.g.* hollow spherical shells of varying icosahedral symmetries.

**RESULTS AND DISCUSSION**

The subunit we employ is a rigid three-dimensional equilateral triangular block made by DNA origami, with a rectangular cross-section of 15 nm × 10 nm, formed by construction of a 6 × 4 square lattice of 2.5 nm diameter double-stranded DNA helices and an edge length of 52 nm [Figure 1D]. The triangular shape is chosen for its mechanical ability to resist shear in the plane of the triangle and to topologically suppress net twist along the perimeter of the triangle because the origami forms a closed loop. The triangular shape is maintained rigid by making the cross-section thick. All information necessary to self-assemble these subunits into user-prescribed high-order structures is encoded in the variable angle modules and bond modules. Fabrication of the core module is similar to our prior work,[19–21] but different design principles are employed for the angle and bond modules as detailed below.

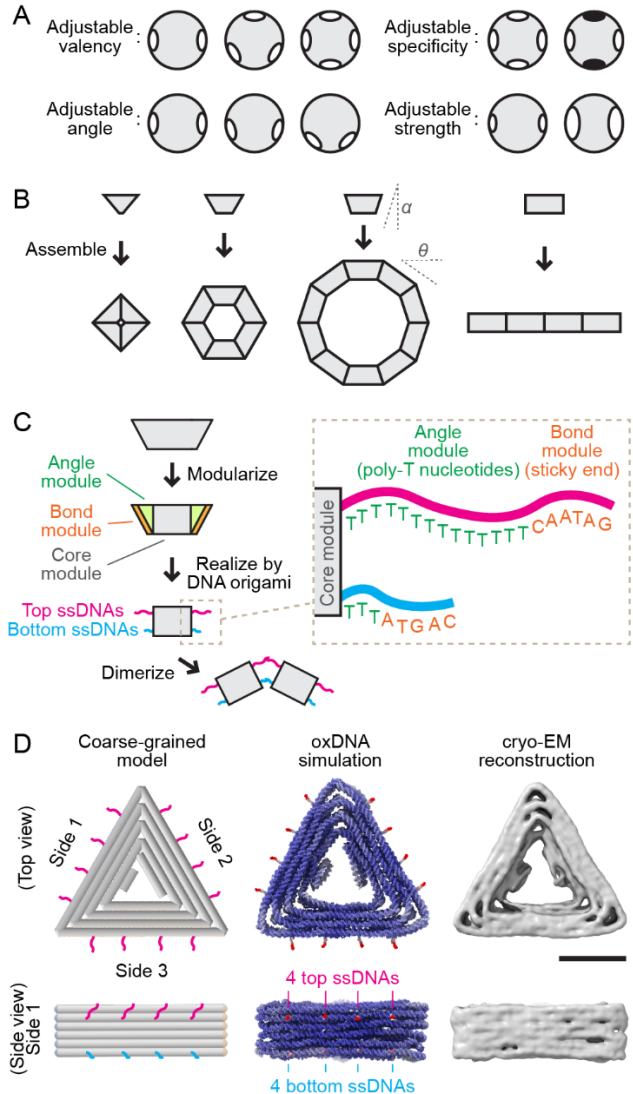

**Figure 1.** Design principles of modular subunits. (A) Schematic illustrating adjustable attributes of patchy particles – valency (*upper left*), bond angle (*lower left*), bond specificity (*upper right*) and bond strength (*lower right*). The white/black patches on particles represent binding sites, with like-colors binding to each other and no interaction between unlike-colors. (B) Subunits with a designed bevel angle $\alpha$ specify the binding angle $\theta = 2\alpha$. (C) Illustration of applying modularity to a subunit block, dividing it into a conserved core module and variable angle and bond modules. On each core face, the variable modules are made of four top (pink) and four bottom (cyan) ssDNA strands. The angle module consists of ssDNA poly-T segments (green sequences); the bond module consists of ssDNA sticky end segments (orange sequences). (D) *Left*: Coarse-grained schematic of the subunit; the variable modules made of ssDNAs are also sketched. Each cylinder represents a double-stranded DNA helix. *Middle*: The oxDNA simulated subunit with color-coded structural rigidity. The core module is mostly rigid (blue); the angle and bond modules are very flexible (red). *Right*: cryo-EM reconstructed subunit core module. Scale bar: 20 nm.

**Design principle and fabrication of subunits**

Subunit geometries define the size and shape of structures they form [Figure 1B].[19–21,30–35] This concept applies to systems ranging from meter-scaled architectures to molecular-level micelle formations. In the former case, wedged-shaped stones are used to build arches. In the latter case, the molecular packing parameter, or ratio of the effective diameters of the hydrophilic and hydrophobic blocks, determines whether spherical, cylindrical or planar micellar assemblies are formed.[36,37]

To reduce the burden of redesigning subunits for different targets,[19–21,29] we propose a design principle that decouples geometry and interactions from the subunit body [Figure 1C]. A *universal* core module is fabricated as a pegboard. Along each outer face of the core, *replaceable* angle and bond modules with encoded geometry and binding information are decorated to individualize the properties of each subunit.

The core module is made from DNA origami, and extruding from each of the three triangle core faces are four "top" and four "bottom" ssDNA strands [Figure 1C-D]. Each top and bottom ssDNA strand has three segments. The innermost segment serves as an anchor and contains a specific DNA sequence that hybridizes with the core module to strongly bind inside the core [Figure S1A-B] and does not vary from design to design. The next two overhanging segments are referred to as the angle and bond modules, extend out from the core and vary with each design [Figure 1C right]. Closest to the core is the angle module, consisting of a series of poly-T nucleotides with varying length, which determines the binding angle when two subunits join ($\theta$ in Figure 1B). Next comes the bond module, comprising 5-8 unpaired nucleotides (the so-called 'sticky end'), which binds specifically with a complementary bond module on another subunit via hybridization. Organization of these 8 ssDNA strands ensures subunit-subunit binding with correct orientations and without offset. Consult the Methods section and Figure S1-S3 for more details.

Note that to achieve productive assembly, the binding strengths (affinities) between subunits should be strong enough to ensure thermodynamic stability of the target structures but weak enough to allow thermal annealing of kinetic traps.[20,38–41] Fine tuning of the binding strength is therefore critical and is permitted by parameters that affect the hybridization strength of the bond modules. Tunable experimental parameters include the number of ssDNA strands per face, the number of nucleotides in each bond module, solution ionic strength and the temperature.

**Self-assembled 2D tiling: program binding specificity using subunits with conserved core**

For simplicity, we first limit our targets to two-dimensional (2D) structures, demonstrating how self-assembled sheets with different tiling patterns arise by varying the interaction specificity encoded in the bond modules [Figure 1C, Figure 2 middle column].

In 2D, the binding angle $\theta$ between adjacent subunits is zero degrees. To accomplish this, all 8 ssDNA strands of the angle module are set to the same length, *e.g.* 3 poly-T [Figure 2A, 2B left]. Note that a short poly-T length is preferred over having no poly-T to overcome steric hindrances between the DNA origami core and to relieve stress within the structure.[42–44]

When the interactions between subunits are designed to have three-fold symmetry, each of the three subunit edges (labeled as S1, S2, S3) have identical bond module sequences and therefore pair with any edge of another subunit [Figure 2A middle]. This is the simplest interaction matrix possible and leads to a 2D sheet with continuous tiling without orientational order, with a single triangle as the fundamental building block [Figure 2A right, see Methods for detailed design].

More complicated tiling patterns can be realized by assigning specific interaction rules between subunits to create directional bonds.[45] As a demonstration, we create a tiling in which the repeating unit is a tetramer [Figure 2B, see Methods for detailed design]. Two 'flavors' of subunits with a 3:1 number ratio are employed to construct a triangularly shaped tetramer [Figure 2B right inset], which join together without orientational order to make the tiling. In contrast to the prior case of a 2D lattice in which each triangle has 3-fold symmetry and there was only one type of bond [Figure 2A middle], here there are two types of bonds [Figure 2B middle]. Except for the bond modules, the two flavors of subunits have the same angle modules and the same core module.

**Self-assembled 3D structures: program local curvature using subunits with conserved core**

Next, we vary the angle module poly-T lengths to demonstrate the ability to assemble three-dimensional (3D) structures with various Gaussian curvatures. The binding angle between two subunits is controlled by the length differences between the four top ssDNA strands and the four bottom ssDNA strands on each face [Figure 1C-D]. In practice, we vary the number of poly-T nucleotides in the angle module, denoted as $\ell_{top}$ and $\ell_{bottom}$ respectively, and keep the same number of bases in the bond module. We expect a zero binding angle if $\ell_{top} = \ell_{bottom}$, a positive binding angle if $\ell_{top} > \ell_{bottom}$, and a negative binding angle if $\ell_{top} < \ell_{bottom}$. These binding angles control the local curvature, thereby defining the final target structure.

Here, we target the creation of icosahedral shells (capsids) and cylinders. The Caspar-Klug theory[2] explains how to transform a 2D plane of triangles into capsids and cylinders by cutting portions of the 2D tiling and wrapping the edges together. We showcase several representative examples as follows.

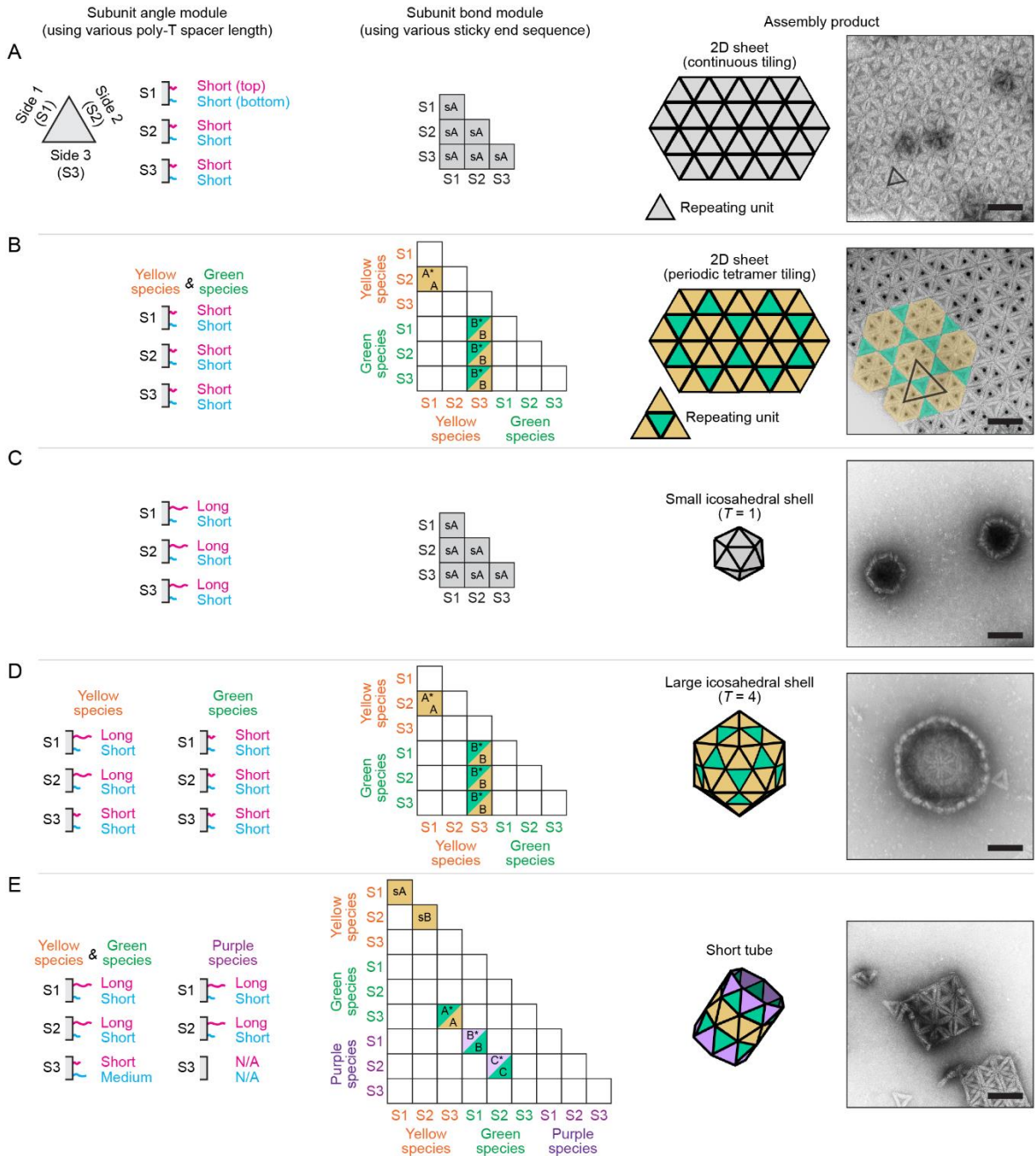

**Figure 2.** Structures self-assembled from subunits with the same conserved core. (A-E) Subunits deploying the same core module, but different angle and bond modules self-assemble into (A) planar sheets with continuous tiling, (B) planar sheets with tetramer tiling, (C) $T=1$ capsid shells, (D) $T=4$ capsid shells, and (E) short tubes. A schematic illustrating the angle modules is provided for each case. Short, medium, and long ssDNA contains a 3, 8, and 14 poly-T segment, respectively (*left column*). The interaction matrix encoded into the bond modules is also stated. Matrix elements with a single color represent a bond between identical species; elements with two colors represent bonds between two different species. The black letters indicate distinct bonds with sequences detailed in Methods (*middle column*). The sketches and TEM images show the final assembled structures. Scale bar: 100 nm (*right column*). Note, the yellow-coded subunits in (B) are labeled with gold nanoparticles to be distinguishable under TEM.

The first case is the icosahedron, which is assembled from 20 equilateral triangles in which all sides can bind to each other with an interaction matrix identical to the prior case of a planar 2D sheet [Figure 2A middle, Figure 2C middle]. However, for the icosahedron, the triangles are designed to join with a positive binding angle of 41.8° by programming all the subunit angle modules to have $\ell_{top}$ = 14 poly-T and $\ell_{bottom}$ = 3 poly-T [Figure 2C left]. The value of $\ell_{top}$ was chosen through systematic trial [see guidance in Figure S4A]. The assembled structures were assessed with transmission electron microscopy (TEM) and gel electrophoresis. The shell comprises 20 identical subunits and is a *T*=1 capsid as classified by the Caspar-Klug theory,[2] in which the triangulation number, *T*, specifies the minimum number of distinct local symmetries/interactions required. [Figure 2C right, Methods]. The design gives very high target specificity and there are no byproducts (other closed structures than *T*=1 capsids) that can be experimentally identified. The overall yield is 62%; that is, 62% of input subunits assemble into target *T*=1 shell, with the rest either remaining as small oligomers or aggregates [Figure S5A].

Next, we create a *T*=4 icosahedral shell [Figure 2D] that shares the same binding rules as the 2D planar tiling [Figure 2B]. Here, as in the planar case, two subunit species are employed [Figure 2D left and middle, Methods]. A flat 'tetramer triangle' composed of 4 subunits are designed by choosing angle modules with $\ell_{top}$ = $\ell_{bottom}$ = 3 poly-T along the three-fold (yellow-green) bond with the expectation that the result would be a 0° binding angle. Twenty such tetrameric triangles are then joined, but with angle modules designed with $\ell_{top}$ = 14 poly-T > $\ell_{bottom}$ = 3 poly-T to make a 41.8° structural binding angle along the five-fold (yellow-yellow) bond to assemble a *T*=4 shell. For the *T*=4 structure there are two distinct bonds, hence four distinct bond modules. One subunit species (green) has identical angle and bond modules on all three sides, while the other (yellow) has two different angle modules and three different bond modules. This $T$=4 structure, with two distinct angle modules and four distinct bond modules, exploits the modularity of our approach. The structures were assessed with TEM and gel electrophoresis, with an overall yield of 34% and again a very high target specificity (no observed byproducts, *i.e.*, no other closed structures than $T$=4 capsids can be experimentally identified) [Figure S5B].

As further validation that this design principle is viable for constructing a variety of complex structures, we demonstrate assembly of short cylindrical tubes, or 'tubelets' [Figure 2E]. Cylinders have zero Gaussian curvature, so to build a cylindrical tube from equilateral triangles requires both positive and negative binding angles.[21] In one of the working designs, we make two core faces with angle modules providing a positive binding angle ($\ell_{top}$ = 14 poly-T > $\ell_{bottom}$ = 3) and one core face with an angle module providing a negative binding angle ($\ell_{top}$ = 3 poly-T < $\ell_{bottom}$ = 8) along the direction perpendicular to and parallel to the axis of symmetry, respectively [Figure 2E left]. An interaction matrix with translational symmetry is deployed [Figure 2E middle, Methods]. The cylinders that formed are monodisperse in length but variable in diameter [Figure S5C] due to the large flexibility between subunits (see the following section).[46–48] This polymorphism, however, could be reduced by increasing the number of distinct bonds in the tubelet, as demonstrated previously.[26,29]

**Characterize the mechanical properties of joints between adjacent subunits**

The joints between subunits are found to be flexible. When freely suspended in solution, thermal fluctuations cause the binding angle between two bound rigid subunits to fluctuate over a broad range, rather than having a single binding angle [Figure 3A top sketch]. This fluctuation is likely a result of employing ssDNA strands in the bonds between paired subunits, as at least 3 bases and up to 14 bases of single-stranded poly-T are present in the variable angle modules [Figure 1C]. Due to the very short persistence length of ssDNA,[49] long strands in the angle modules will coil, leading to spring-like joints that bend, twist and stretch.

To enable quantitative determination of the flexibility of the joint, we use cryo-EM and multi-body refinement[24] to measure the distribution of binding angles between subunits in a dimer [Figure 3]. We chose to measure dimer fluctuations because monomer addition is the fundamental step in assembly and because dimer fluctuations are the simplest to study experimentally. Importantly, because the origami core module is designed to be much more rigid than the joint (the angle and bond modules), the molecular motion of a swinging dimer is described by two rigid subunit cores that maintain their internal structures but differ in their relative orientations.

Briefly, multi-body refinement builds on focused refinement using the partial signal subtraction technique.[50–52] By selectively masking away electron density corresponding to the flexible complex outside of the user-defined part, *i.e.,* a rigid subunit core, reconstructions with higher resolution can be achieved. During the refinement process, the relative orientation of the two independently-moving units is determined for every individual dimer image. The principal component analysis of the refined relative orientations leads to extraction of the principal motions. Structural motions of the dimer can then be expressed by linear combinations of these orthogonal principal motions. See Methods for more technical details.

Note, the distribution of dimer angles derived directly from cryo-EM data is expected to deviate from the room-temperature distribution. Despite rapid cooling rates of $10^6$ K/sec[53–56], water vitrification takes approximately $10^{-4}$ seconds,[57,58] allowing sufficient time for the binding angle distribution to thermally equilibrate, given the rotational diffusion time constant of $10^{-9}$ seconds for a 50 nm subunit.[59] In our analysis, we assume the cryo-measured angle distribution is equilibrated at 136 K and rescale the Boltzmann factor to 298 K to determine the room-temperature angle distribution [Methods]. Other temperature-dependent effects, such as potential changes in ssDNA persistence length,[60] are not considered in this analysis.

The cryo-EM images of Figure 3A underpin our physical intuition in regard to the long ssDNA ($\ell_{top}$ = 14 poly-T) acting as an entropic spring and the short ssDNA ($\ell_{bottom}$ = 3 poly-T) acting as a hinge. The relative angle distribution between the two subunits is measured from the cryo-EM images to have a Gaussian distribution with an average binding angle of 21.6° and a standard deviation of 12.7° after temperature rescaling [Figure 3A bottom, Figure S6A, Video S1]. An estimated room-temperature bending elastic modulus 20.5 $k_BT/rad^2$ of the joint is extracted from the distribution of binding angles. We reiterate that we assume that the physical properties of the DNA are temperature independent and thus we assume that the average binding angle at the vitrification temperature and the elastic modulus is the same as at room temperature.

As a comparison, when all ssDNA strands in the angle modules are short, with $\ell_{top} = \ell_{bottom}$ = 3 poly-T, the joint is more rigid [Figure 3B top], with a narrower Gaussian distribution with a standard deviation of 7.4° and an average binding angle of 4.6° at room temperature [Figure 3B bottom, Figure S6B, Video S1]. This leads to a corresponding larger room-temperature bending elastic modulus of 59.3 $k_BT/rad^2$.

Finally, we note that besides the dominant bending motion, minor twisting and stretching modes are also observed, revealing that it is oversimplified to view the bottom row of ssDNA as a hinge with only one degree of freedom [Figure S7].

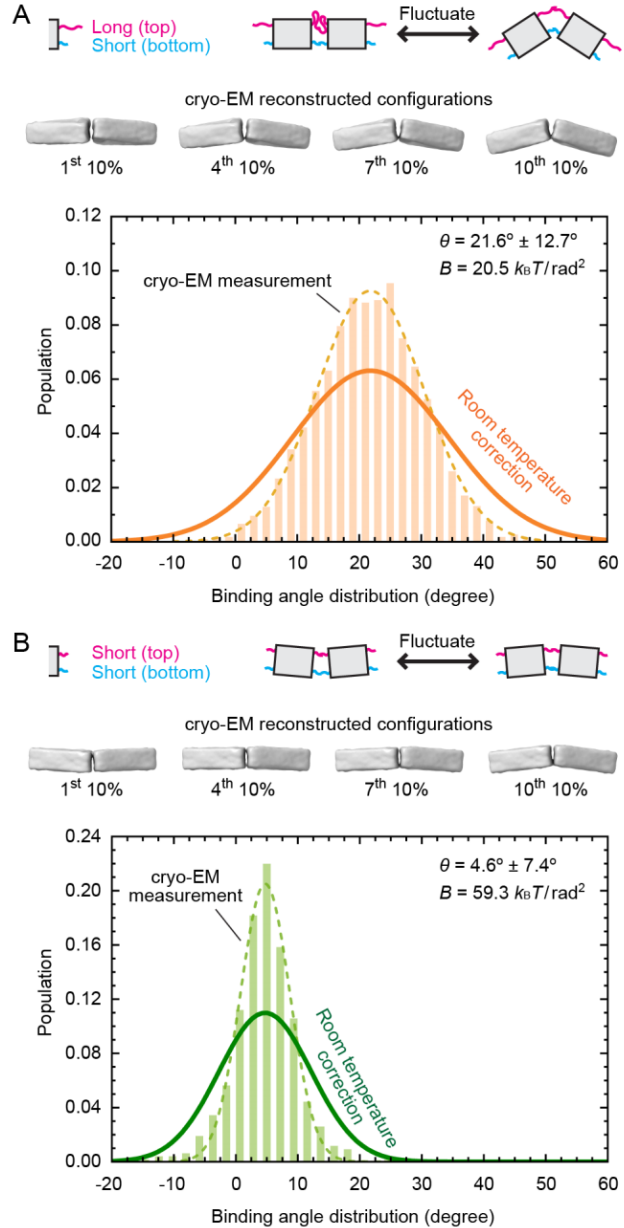

**Figure 3.** Angular distribution between subunits characterized by cryo-EM. (A) The dimerized subunit pair with angle modules $\ell_{\text{top}} = 14$ poly-T $> \ell_{\text{bottom}} = 3$ poly-T (used in Figure 2C and Figure 2D yellow-yellow bonds) has a broad angular distribution. The four cryo-EM reconstructions represent average configurations of the most flattened 10% of the distribution, the most bent 10% ensemble, and another two states in between (*top*). The binding angle distribution of the joint is extracted directly from cryo-EM observation (bar graph, 136 K), fitted by a Gaussian (dashed curve), and rescaled to room temperature (solid curve, 298 K) (*bottom*). (B) The dimer joined by angle modules $\ell_{\text{top}} = \ell_{\text{bottom}} = 3$ poly-T is more rigid compared to (A), visualized by smaller changes in configurations (*top*) and a narrower binding angle distribution (*bottom*).

**Flexible joints between subunits as a feature**

Next, we connect the mechanical properties of the flexible subunit-subunit joints to the larger-scale assemblies that form, and we discuss the advantages offered by the flexibility.

As shown in Figure 2C and Figure 2D, the angle module $\ell_{top}$ = 14 poly-T > $\ell_{bottom}$ = 3 poly-T gives required five-fold vertices in either *T*=1 or *T*=4 capsid assemblies, with structural binding angles of 41.8°. Although the corresponding (temperature-corrected) cryo-EM measurement shows a dimer fluctuation with an average angle of only 21.6°, the required 41.8° angle can still form because the dimer flexibility is great enough for thermal fluctuations to open the dimer [Figure 3A, Figure 4A left]. The dimer, however, is not flexible enough to allow other closed forms (*e.g.* an octahedron with a binding angle of 70.7°), explaining the high specificity and yield of assembly [Figure S5A-B]. Note that the fluctuating dimer also reaches 0° [Figure 3A, Figure 4A left], but we never observe planar structures. We hypothesize that icosahedra, with 20 subunits, are observed rather than large flat sheets, because closed structures are kinetically and energetically favored over large ones. Once the closed icosahedral structure forms, it is more stable than an open sheet because all subunits in an icosahedron have three neighbors, while subunits along the edge of an open sheet have less than three neighbors and thus open structures dissolve more readily than closed ones.

Our results suggest that flexible joints lead to assembly fidelity as long as two requirements are satisfied when designing subunits for a certain target structure. The first is energetic; (1) the fluctuation range of the individual subunit-subunit bond angle should be large enough to cover the target binding angle. The second is kinetic; (2) within the fluctuation range, the target binding angle should be the largest among other binding angles that lead to closed structures other than the target [Figure S8A]. The second criterion is because if there is more than one possible polymorph within the fluctuation range, the closed structure with the smallest number of subunits (largest binding angle) is kinetically favored [Figure S8B].[44] Note that in this system, wherein the energy cost of bending is relatively low, a closed structure is energetically favorable over an open structure with the same number of subunits due to lower edge tensions.[61–64] Rather than restricting subunits to join with others with precise angles, which has a large entropic penalty and thus we hypothesize will lower the rate of assembly, these design principles advantageously offer larger error tolerance in design and fabrication as well as higher assembly rates.

In regard to the yield of a target structure, flexibility and complexity can be traded off, meaning that if high flexibility leads to polymorphism, then specificity can be restored by increasing the number of distinct subunits.[26–29] For example, the angle module, $\ell_{top}$ = 14 poly-T > $\ell_{bottom}$ = 3 poly-T [Figure 3A], can be programmed to preferentially form different structural binding angles [Figure 4A] by increasing the number of distinct subunits from one to two. Consider the case of one subunit species, in which any two triangular subunits can pair to form a bond. With this binding rule, the flexible joints favor forming the 41.8° binding angle that makes a five-fold vertex [Figure 4A bottom right]. By contrast, a 0° binding angle and six-fold vertex are preferred I if two distinct subunit species are required to form a dimer. In this case, odd-numbered clusters are suppressed [Figure 4A top right]. Depending on the assigned interaction matrix, different vertexes and local structural curvatures can therefore be selectively enriched [Figure S9].

Last, we show that assemblies of higher-order structures can be targeted with flexible bonds, by using the same angle module to achieve widely different binding angles by controlling bond specificity using the bond modules. We demonstrate construction of a *T*=4 icosahedral shell as proof of concept by using the same angle module, $\ell_{top}$ = 14 poly-T > $\ell_{bottom}$ = 3 poly-T, for all its joints [Figure 4B]. Here, both the required 0° and 41.8° binding angles (along the yellow-green and yellow-yellow bond shown in Figure 4B, respectively) are realized by encoding the binding rules described in Figure 4A to bias the angle degeneracy. The resultant assembly has high product specificity (no experimentally-identified off-target closed structures). We note that an even higher yield can be achieved if different and more appropriate angle submodules were used, but our point here was to demonstrate that by increasing flexibility, a range of specific and widely different binding angles can be selected by increasing the number of distinct bond modules.

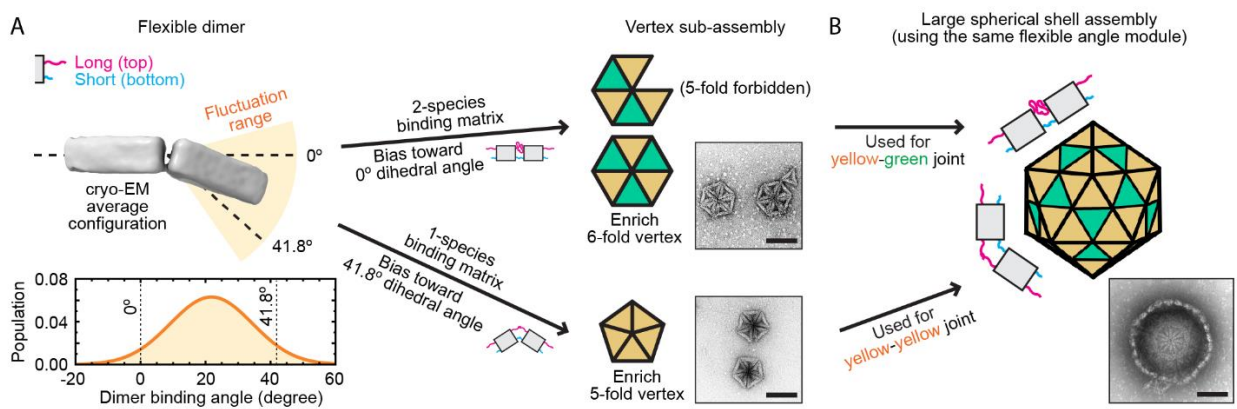

**Figure 4.** Flexible subunit-subunit joints allow formation of bonds with different binding angles. (A) By assigning an appropriate interaction matrix via bond modules, sub-assembly of a 5-fold vertex (*bottom right*, corresponding to a 41.8° binding angle) or a 6-fold vertex (*top right*, corresponding to a 0° binding angle) can be preferentially enriched using subunit-subunit joints with the same flexible angle module (*left*, characterized in Figure 3A). (B) Following the same principle, a *T*=4 large capsid shell can be constructed using the same angle module for all its joints to target the required 0° and 41.8° binding angles. Scale bar: 100 nm.

**CONCLUSIONS**

We applied the engineering notion of modularity to nano-scaled subunits, enabling construction of user-prescribed large and complex self-assembled structures in an economical manner. In this design strategy, each subunit contains one triangular core module and a set of angle and bond modules, composed of 8 overhanging ssDNA strands, on each of the three core faces. The angle and bond modules are largely independently tunable and control the binding angle and the binding strength/specificity, respectively [see Methods and Figure S3], while the core module is a non-functional block conserved among all variants.

Compared to prior approaches,[19–21,29,32] this modular principle dramatically reduces the design and synthesis effort, and subsequently the cost, of subunit variants. The total 24 ssDNA strands in the angle/bond modules (out of 204 strands in a single subunit) are the only parts that need to be re-designed for each variant, and their design relies on simple geometrical and hybridization rules instead of complex origami engineering. The angle module contains poly-T segments; the only design consideration is their length [Figure S1]. The bond module uniquely pairs with a second bond module on another subunit; the most important design consideration is its complementary hybridization sequences [Figure S1C, Figure S2]. Our bond module design and assembly conditions also optimize the kinetics by ensuring a high off-rate to favor equilibration in a short time, flexibility to ensure a high on-rate, and a free energy of binding that is a few times thermal energy. This scheme eliminates the most labor-intensive process of designing a universal core module[65] and therefore makes the platform suitable for researchers with minimal training in DNA origami. Additionally, it is straightforward to extend other oligos from the core on the upper and bottom portions [Figure S1, SI Sequence] from which one can attach nanoparticles and/or biomolecules for introducing a range of functionalities.[66–70]

Besides providing an economical approach to assemble complex structures, we introduced a new method quantifying the degree of flexibility of dimerized subunit-subunit pairs using cryo-EM.[24] We directly visualized the probability distribution of bending angles between adjacent subunits, from which we extracted the bending elastic modulus. This information offers a straightforward perspective on how flexibility contributes to the assembly of high-order structures. Looking forward, the impact of the cryo treatment on the measurement of the elastic modulus is not fully understood. The changes in Young's modulus and persistence length of individual DNA strands during vitrification, and how these affect the measured angular distribution, remain unknown and warrant further investigation.

Our findings suggest that incorporating a judicious amount of flexibility in the bond offers benefits of error tolerances in design and fabrication, while still guaranteeing assemblies with high yields and high specificity of the target structure. Of course, too much flexibility often leads to the production of assemblies other than the design targets. However, we show here that fidelity to the designed assembly can be restored by introducing more distinct subunit species [Figure S9].[26–29]

We demonstrated the versatility of the modular concept by engineering several self-assembled structures with different Gaussian curvatures. This success suggests that the method is general and will apply to the wide variety of self-assembled structures imagined by the readers. While we choose to realize the concept using DNA origami, this modular engineering principle could also be applied more broadly.[71–73]

**METHODS**

**Design of the DNA origami subunits.** The subunit core module, made from DNA origami, was designed using caDNAno v0.2 [74] [Figure S1A], based on multilayer concepts[23,75] and folded through a one-pot reaction procedure.[76] The design uses 204 synthetic short ssDNA oligos (staples) to hybridize with and to 'fold' one long single-stranded circular DNA (scaffold) into the target structure. The sequences for all core oligos can be found in the SI Sequence file. 24 out of the total 204 ssDNA oligo strands [Figure S1A pink- and cyan-labeled strands; exemplary sequences in Figure S1B] were extended, adding poly-T nucleotides segment (the angle module) and 'sticky end' segment (the bond module) to encode geometry and binding information for assemblies, respectively [Figure 1C right]. These 24 strands (4 top and 4 bottom strands on each of the three core faces [Figure 1D]) are the only part that needs to be modified to make distinct subunit species.

To ensure the binding between subunits has the correct orientation, sequences of the 4 top strands were designed to be different from those of the 4 bottom strands. Additionally, the arrangement of the strands along a row establishes a chirality to each triangle, e.g. a particular row might have the binding sequence a, b, b*, a* with the same sequence repeated on each face [Figure S1C, Figure S2]. The symmetry-breaking and chirality prevent flipping of subunits, thereby defining the inside/outside surface when assembled into closed structures. To avoid having the faces bind with an offset, all 4 top strands (and all 4 bottom strands) were assigned with different sequences so that binding in any configurations different from the designed one will be weak and therefore transitory while the correct configuration will be strong and long lasting.

Selection of the number of binding strands per subunit face and the number of hybridization pairs (sticky end length) per strand was set by several criteria. (1) The total number of sites to extend ssDNA from the core was limited, with 4 per row being near the maximum. (2) One wants to maximize the off-rate of subunit unbinding, which allows monomers that bind in incorrect positions to fall off, re-bind, and equilibrate the assembly into the designed structure. Maximizing the off-rate, which is independent of subunit concentration, is done by reducing the binding strength by decreasing the number of bases involved in hybridization bond module. (3) One can only reduce the binding strength so much as it is necessary for the binding free energy to exceed $k_BT$ for assembly. The binding free energy is a monotonic function of the ratio of the on- to off-rate constant, so if we fix the binding free energy to be several $k_BT$, then we must compensate by increasing the on-rate as we increase the off-rate. We do so by increasing the monomer concentration as much as practical. Due to cost constraints and other factors such as suspension stability, the optimal concentration is in the range of 5 – 100 nM. In summary, our design rules are to maximize the number of strands per face, minimize the number of hybridized base pairs per strand and maximize the monomer concentration, while maintaining the binding free energy near several $k_BT$.

In this study we put binding strands on the 2$^{nd}$ and the 6$^{th}$ rows of dsDNA comprising the subunit face [Figure 2D, Figure S1C], as we employ the binding strands to set the binding angles and two points determine a straight line. With 2 rows of 4 binding strands per subunit face, we found that assembly was optimized with 5-8 binding base pairs per strand [Figure S4B]. We also found that there was a weak correlation between the number of poly-T segment length in the angle module and the needed number of hybridized bases in the bond module. The number of binding base pairs per strand needs to be increased with the poly-T base count [Figure S3, Figure S4A], approximately by an average of 0.5 binding base pairs for every 5 poly-T nucleotides added. (We suggest using 5 binding base pairs for $\ell_{bottom}$ = 3 poly-T; 5 binding base pairs for $\ell_{top}$ < 10 poly-T and 6 binding base pairs for $\ell_{top}$ = 10-14 poly-T [Figure S3].) We hypothesize this was due to the binding on-rate decreasing with long angle modules because of the entropic cost for the poly-T region to move away from the bases involved in the bond module. The important design principle is that the bond and angle modules are not fully independent, although the coupling is weak.

**Folding and purification of the DNA origami subunits.** The folding mixtures contain 50 nM p8064 scaffold (Tilibit Nanosystems), staple oligonucleotides (Integrated DNA Technologies) of 200 nM each, and a folding buffer. The buffer contains 5 mM Tris base, 1 mM ethylenediaminetetraacetic acid (EDTA), 5 mM sodium chloride (NaCl), and 15 mM magnesium chloride ($MgCl_2$) (Sigma-Aldrich). The folding mixtures were then subjected to a thermal annealing ramp (80 °C for 2 minutes, 65 °C for 15 minutes, then cooling with a 1 °C/hour rate from 58 °C to 50 °C) in a thermal cycling device (Bio-Rad Laboratories).

All folded subunits were then gel purified (to remove excess oligonucleotide strands and misfolded aggregates) and concentrated (using ultrafiltration, 100 kDa molecular weight cutoff Amicon Ultra Centrifugal Filter Unit) before being used for assembly experiments. An exemplary purification agarose gel is shown in Figure S1D, wherein the monomer 'band' contains the target species. A NanoDrop microvolume spectrophotometer (Thermo Fisher Scientific) was used to check the subunit concentrations. Both procedures were performed following details previously described.[65]

**Assembly of subunits into desired structures.** All self-assembly experiments were conducted with a total subunit concentration of 5 nM. The assembly solutions contain 5 mM Tris base, 1 mM EDTA, 5 mM NaCl, and 20 mM $MgCl_2$. The samples were then incubated in a thermal cycling device at 40 °C for one hour and quenched at roughly 6 °C/min to the target assembly temperature for a certain reaction time. Optimal assembly conditions vary from structure to structure.

Here, we provide general strategies to efficiently pick optimal parameters when exploring new designs for desired target structures. (1) *Assembly solution and temperature*. Assembly depends strongly on temperature and relatively weaker on solution magnesium concentration. We suggest beginning with 20±2.5 mM MgCl$_2$ and screening assembly temperature within the range of 20-40 °C. As articulated in the main text, we aim for the highest off rate possible. We find that the optimal assembly typically occurs 1-2 °C below the structure melting temperature, at which binding strengths between subunits are just strong enough to trigger formations of high-order structures. This can be easily evaluated by negative stain EM [Figure S4C]. Note, both lower assembly temperature and higher MgCl$_2$ concentration gives stronger subunit-subunit binding. (2) *Bond modules*. As stated in the "Design the DNA origami subunits" section, we suggest using bond modules with 5-8 binding base pairs per strand [Figure S3]. Again, optimal assembly happens when the binding strength is just strong enough for formations of high-order structures. (3) *Angle modules*. To determine a proper angle module, we suggest fixing $\ell_{bottom}$ = 3 poly-T and varying $\ell_{top}$ when targeting a positive binding angle, and fixing $\ell_{top}$ = 3 poly-T and varying $\ell_{bottom}$ when targeting a negative binding angle [Figure S4A]. To reduce the effort, first evaluate vertex subassembly when varying $\ell_{top}$ or $\ell_{bottom}$, before checking the assembly of whole structures. For example, a 6-fold (hexamer), 5-fold (pentamer), and 4-fold (tetramer) vertex is expected for binding angle of 0°, 41.8°, and 70.7°, respectively. A binding angle between these numbers would yield a mixture of neighboring close structures.

**Assembly: 2D sheets with continuous tiling pattern.** To construct a 2D sheet with continuous tiling without orientational order, we designed one subunit species with isotropic interaction matrix. For individual subunits, identical angle modules and bond modules are applied to all three triangular faces S1, S2, and S3 of the core module [Figure 2A]. The angle modules employ poly-T segment $\ell_{top} = \ell_{bottom} = 3$ poly-T [Figure S1B, Figure S3]. The binding modules employ 5 bps sticky end segment with 'sA' sequence detailed in Figure S2. The assembly temperature is 28 °C with an assembly time of 36 hours.

**Assembly: 2D sheets with periodic tetramer tiling.** Two subunit species were designed for this case, color-coded as 'yellow' and 'green' in Figure 2B, and were mixed with a 3:1 number ratio. The angle modules of both species, for all S1, S2, and S3, employ $\ell_{top} = \ell_{bottom} = 3$ poly-T [Figure S1B, Figure S3]. The 'yellow' species binding modules employ 5 bps 'A' sequence for S1, 'A*' sequence for S2, and 'B' sequence for S3 [Figure S2]. The 'green' species binding modules employ 5 bps 'B*' sequence for all S1, S2, and S3. The assembly temperature is 28 °C with an assembly time of 36 hours.

**Assembly: 3D small spherical shells ($T$=1 capsid).** One subunit species was designed, with identical angle module and bond module for all its S1, S2, and S3 [Figure 2C]. The angle modules employ $\ell_{top} = 14$ poly-T > $\ell_{bottom} = 3$ poly-T [Figure S1B, Figure S3]. The binding modules employ 'sA' sequence (top 6 bps, bottom 5 bps) [Figure S2]. Note, the binding matrix used here is the same as the one for '2D sheets with continuous tiling pattern'. The assembly temperature is 25 °C with an assembly time of 24 hours.

**Assembly: 3D large spherical shells ($T$=4 capsid).** Two subunit species were designed, color-coded as 'yellow' and 'green' in Figure 2D, and were mixed with a 3:1 number ratio. For the 'yellow' species, angle modules employ $\ell_{top} = 14$ poly-T > $\ell_{bottom} = 3$ poly-T for S1 and S2, and $\ell_{top} = \ell_{bottom} = 3$ poly-T for S3 [Figure S1B, Figure S3]; the bond modules employ 'A' sequence (top 6 bps, bottom 5 bps) for S1, 'A*' sequence (top 6 bps, bottom 5 bps) for S2, and 7 bps 'B' sequence for S3 [Figure S2]. For the 'green' species, angle modules employ $\ell_{top} = \ell_{bottom} = 3$ poly-T for all S1, S2, and S3 [Figure S1B, Figure S3]; the bond modules employ 7 bps 'B*' sequence for all S1, S2, and S3 [Figure S2]. Here, the yellow-green (B-B*) bond was designed to be stronger than the yellow-yellow (A-A*) bond to induce hierarchical assembly,[20] biasing formation of tetramer subassemblies. Note, the binding matrix used here is the same as the one for '2D sheets with periodic tetramer tiling'. The assembly temperature is 30 °C with an assembly time of 96 hours.

**Assembly: short tubes.** To construct 3-layer short tubes, we designed three subunit species, color-coded as 'yellow', 'green', and 'purple' in Figure 2E, and were mixed with a 1:1:1 number ratio. The angle modules of all three species employ $\ell_{top}$ = 14 poly-T > $\ell_{bottom}$ = 3 poly-T for S1 and S2, and $\ell_{top}$ = 3 poly-T < $\ell_{bottom}$ = 8 poly-T for S3, except the 'purple' species whose S3 has no angle modules attached [Figure S1B, Figure S3]. The 'yellow' species bond modules employ 'sA' sequence (top 6 bps, bottom 5 bps) for S1, 'sB' sequence (top 6 bps, bottom 5 bps) for S2, and 5 bps 'A' sequence for S3. The 'green' species bond modules employ 'B' sequence (top 6 bps, bottom 5 bps) for S1, 'C' sequence (top 6 bps, bottom 5 bps) for S2, and 5 bps 'A*' sequence for S3. The 'purple' species bond modules employ 'B*' sequence (top 6 bps, bottom 5 bps) for S1, 'C*' sequence (top 6 bps, bottom 5 bps) for S2, and no bond modules for S3 [Figure S2]. The assembly temperature is 30 °C with an assembly time of 96 hours.

**Conjugate gold nanoparticles to DNA origami subunit.** The gold nanoparticles (AuNPs, Ted Pella), 10 nm in diameter, were first functionalized with thiol-modified ssDNA (5'-HS-C$_6$H$_{12}$-TTTTTAACCATTCTCTTCCT-3', Integrated DNA Technologies) following scheme described in Ref. 77. The targeting subunits were also labeled with ssDNA handles with complementary sequence (5'-AGGAAGAGAATGGTT-3') on their interior edges, following descriptions detailed in Ref. 45. We first assembled subunits into desired high-order structures using optimal assembly conditions. The sample solution was then mixed with the AuNP suspension, with a final particle concentration five times larger than the concentration of targeting subunits. The mixture was incubated for 12 hours in the native buffer condition before imaging.

**Negative stain electron microscopy.** The samples were first prepared using FCF400-Cu grids (Electron Microscopy Science, glow discharged at -20 mA for 30 seconds at 0.1 mbar using Quorum Emitech K 100X glow discharger before usage) and 2 wt% uranyl formate solution. The images were taken using an FEI Morgagni Transmission Electron Microscope, operated at 80 kV, with a Nanosprint5 complementary metal-oxide semiconductor camera (AMT). Images were acquired between ×8000 and ×22,000 magnification.

**Agarose gel electrophoresis.** The assembly yields were investigated using agarose gel electrophoresis. We employed 0.5 wt% agarose gels containing 0.5x TBE, 3.75 % SYBR-safe DNA gel stain, and 20 mM $MgCl_2$. The gel electrophoresis was performed at 80 V bias voltage at 4 °C, for varying time (2-4 hours) with buffer exchanged every 40 minutes. The gels were then scanned using a Typhoon FLA 9500 laser scanner (GE Healthcare) at a 25 μm resolution [Figure S5].

The intensity profile of each gel lane was extracted with the background signal subtracted (obtained from the profile of an empty lane). The processed intensity profile was then normalized and fitted with a Gaussian to the target structure peak. The area underneath the Gaussian curve was then defined as the yield (fraction) of successful assemblies.

**Cryogenic electron microscopy.** Higher concentrations of DNA origami subunit are used for cryo-EM grids than for assembly experiments. To prepare samples, we typically prepare 1-2 mL folding mixture, gel purify it, and concentrate the sample by ultrafiltration. EM samples are prepared on glow-discharged C-flat 1.2/1.3 400 mesh grids (Protochip). Subunits with a single active bond are prepared and suspended in a buffer containing 5 mM Tris base, 1 mM EDTA, 5 mM NaCl, and 5 mM $MgCl_2$. To ensure that dimers have formed before plunging, subunit solution is mixed 1:1 with 35 mM $MgCl_2$, bringing their salt concentration to 20 mM $MgCl_2$. The solution then sits at room temperature for 30 min. Plunge-freezing of grids in liquid ethane is performed with an FEI Vitrobot VI with sample volumes of 3 µL, wait time of 60 s, blot time of 9 s, and blot force of 0 at 22 °C and 100% humidity.

Cryo-EM images for DNA origami dimers were acquired with a Tecnai F20 TEM with a field emission gun electron source operated at 200 kV and Compustage, equipped with a Gatan Oneview CMOS camera. Particle acquisition is performed with SerialEM. The defocus is set between -1.5 and -4 µm for all acquisitions with a pixel size of 3.757 Angstrom.

**Single-particle reconstruction and multi-body refinement.** Image processing is performed using RELION-4.[78] Contrast-transfer-function (CTF) estimation is performed using CTFFIND4.1.[79] After picking single particles (subunits), we perform a reference-free 2D classification from which the best 2D class averages are selected for processing, estimated by visual inspection. The particles in these 2D class averages are used to calculate an initial 3D model. A single round of 3D classification is used to remove heterogeneous monomers and the remaining particles are used for 3D auto-refinement and post-processing. The post-processed maps are deposited in the Electron Microscopy Data Bank.

Fluctuations of subunits were processed using RELION-4's multi-body refinement.[24] After getting a postprocessed reconstruction of a dimer using single-particle reconstruction, we create masks around the two triangular cores using the eraser tool in ChimeraX.[80] These were used in the '3D multi-body' job in RELION 4 to get the set of fluctuations the two bodies in the dimer. Outputs of the multi-body refinement are the principal components (PCs) of the fluctuations of the two bodies and corresponding density maps for the two bodies for different eigenvalues along the eigenvectors of the PCs. By measuring the binding angles of the dimers with respect to each other in the PC density maps, we are able to relate the PC eigenvalues for a given eigenvector to a binding angle of the dimer. This eigen-space to real-space analysis is shown in Figure S6.

To obtain room-temperature distribution of dimer angles, temperature corrections are required. Water forms a vitreous glass near 136 K, and even though the cryo cooling rate is high, with cooling rates of $10^6$ K/sec[53–56], it still takes about $10^{-4}$ seconds for water to vitrify.[57,58] This is more than enough time for the binding angle distribution to thermally equilibrate given that the rotational diffusion time constant of a 50 nm subunit is $10^{-9}$ seconds.[59] Thus, in our analysis, we assume that the cryo-measured angle distribution is equilibrated at 136 K and so we rescale the Boltzmann factor from 136 K to room temperature at 298 K to determine the angle distribution at room temperature. Other temperature effects, such as the possibility of a temperature-dependent ssDNA persistence length,[60] which could cause the average binding angle measured at the cryogenic temperature, as well as the effective spring constant, to differ from the room temperature one, are ignored here.

**ASSOCIATED CONTENT**

**Supporting Information**

The Supporting Information is available free of charge online.

Supplementary figures (Figure S1 – Figure S9)

**Corresponding Author**

Seth Fraden – Martin A. Fisher School of Physics, Brandeis University, Waltham, MA 02453, USA; orcid.org/0000-0002-2420-9939; Email: fraden@brandeis.edu

**Author Contributions**

W.-S.W., W.B.R., and S.F. conceived the idea and designed the research; W.-S.W., T.E.V., D.H., and R.S. performed research; W.-S.W., T.E.V., D.H., R.S., W.B.R., and S.F. worked on different facets of data analysis; W.-S.W. and S.F. wrote the paper, and all authors contributed to the final manuscript.

**Notes**

The authors declare no competing interest.

**ACKNOWLEDGEMENTS**

We thank Pragya Arora, John Berezney, Greg Grason, Yi-Yun Ho, Katsu Nishiyama, Michael Norton, Michael Stehnach, and Quang Tran for helpful discussions, Berith Isaac and Amanda Tiano for their technical support with electron microscopy. TEM images were prepared and imaged at the Brandeis Electron Microscopy Facility. We acknowledge financial support from the Materials Research Science and Engineering Center (MRSEC) at Brandeis University funded by the NSF DMR-2011846.

# Supporting Information

# Economical and versatile subunit design principles for self-assembled DNA origami structures

Wei-Shao Wei[1,2], Thomas E. Videbæk[1,2], Daichi Hayakawa[1,2], Rupam Saha[1,2], W. Benjamin Rogers[1,2], Seth Fraden*[1,2]

[1] Martin A. Fisher School of Physics, Brandeis University, Waltham, MA 02453, USA
[2] Materials Research Science and Engineering Center (MRSEC), Brandeis University, Waltham, MA 02453, USA

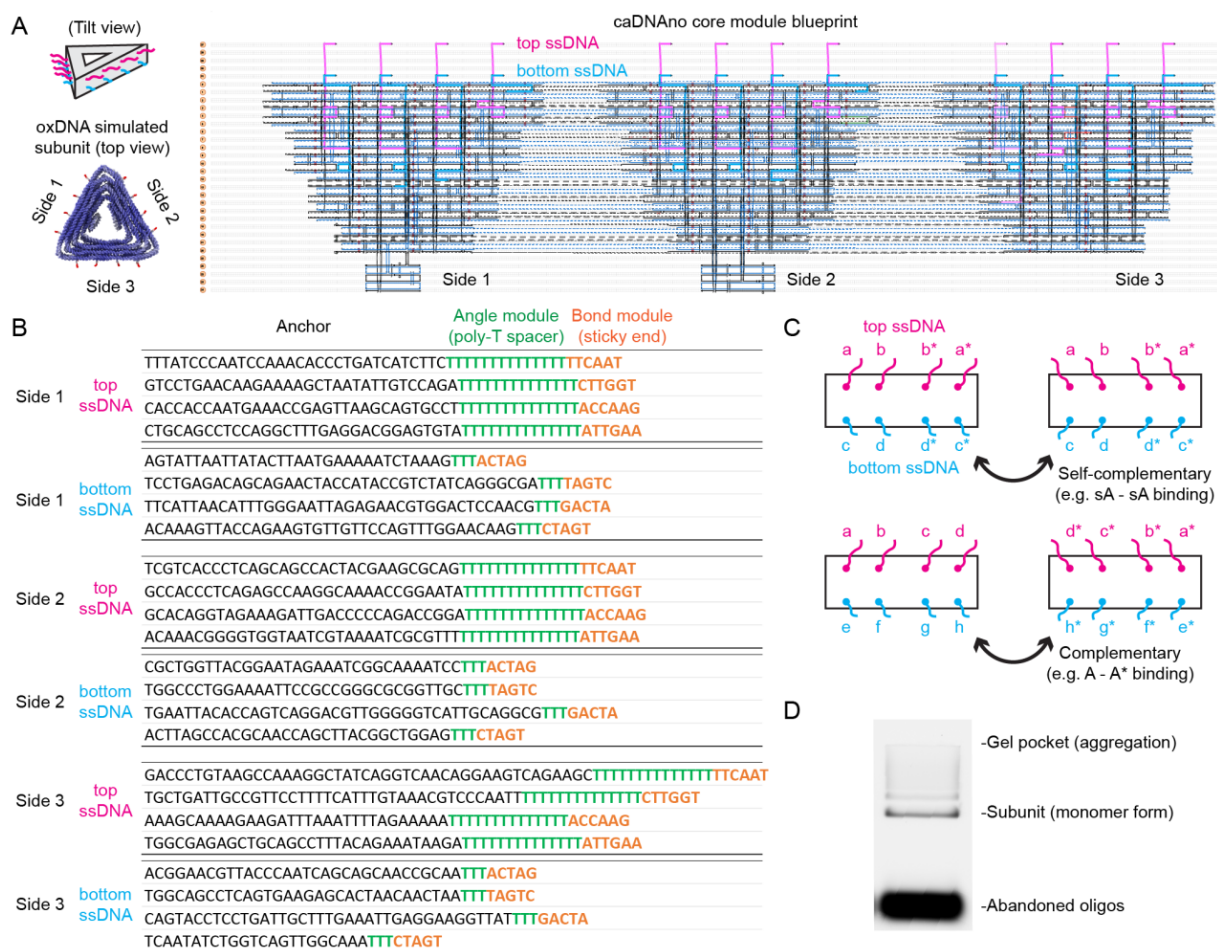

**Figure S1.** Design of the subunit core, angle, and bond module. (A) The core module designed using caDNAno, with one circular p8064 scaffold and 204 oligos (see SI Sequence file for the full oligo sequences). On each subunit triangle face, 4 top (pink) and 4 bottom (cyan) ssDNA strands compose the angle and bond module. Inset: sketch of a tilt-view subunit and an oxDNA-simulated top-view subunit. (B) 5' to 3' sequences of the total 24 variable ssDNAs. Each strand contains 3 segments – conserved 'anchor' (black), variable angle module (poly-T spacer, green), and variable bond module (sticky end, orange). Here, the angle modules employ $\ell_{top}$ = 14 poly-T, $\ell_{bottom}$ = 3 poly-T, the bond modules employ 'sA' sequence (Figure S2, Figure S3) which make $T$=1 capsids shown in Figure 2C. (C) The organization of 8 ssDNAs on each core face for self-complementary binding (*top*) or for binding between two species (complementary; *bottom*). a-a*, b-b*, etc. indicate complementary ssDNA sequences (Figure S2). (D) An exemplary purification agarose gel, with the band labeled 'subunit (monomer form)' contains the target species.

|  |  | 1 | 2 | 3 | 4 |
|---|---|---|---|---|---|
| sA | top | TTCAATT | CTTGGTA | TACCAAG | AATTGAA |
|  | bottom | ACTAGCA | TAGTCTA | TAGACTA | TGCTAGT |
| sB | top | ATAAGT | TCCTTC | GAAGGA | ACTTAT |
|  | bottom | TCAGAC | TACCTT | AAGGTA | GTCTGA |
| sC | top | CCATTC | CTCGAA | TTCGAG | GAATGG |
|  | bottom | AGTTAC | GTCTAG | CTAGAC | GTAACT |
| sD | top | GATCTT | TCCACA | TGTGGA | AAGATC |
|  | bottom | TTAACC | TTGGAT | ATCCAA | GGTTAA |
| A | top | ATGACA | AACCTA | GAGACA | ACTAAC |
|  | bottom | CAATAG | CTAGGA | CACATC | ACCTGA |
| A* | top | GTTAGT | TGTCTC | TAGGTT | TGTCAT |
|  | bottom | TCAGGT | GATGTG | TCCTAG | CTATTG |
| B | top | AGATAGTG | TTCCATAC | GATATGTG | AACATTAT |
|  | bottom | GCATCTCA | AGATTCGA | TTCTCACC | TCGTACTC |
| B* | top | ATAATGTT | CACATATC | GTATGGAA | CACTATCT |
|  | bottom | GAGTACGA | GGTGAGAA | TCGAATCT | TGAGATGC |
| C | top | GGATAA | GGTATT | GGTAAT | AGAGAT |
|  | bottom | AGTTCC | ATTCTG | ATTCAG | GTAGAT |
| C* | top | ATCTCT | ATTACC | AATACC | TTATCC |
|  | bottom | ATCTAC | CTGAAT | CAGAAT | GGAACT |
| D | top | GAGACA | GACAGA | TACAGG | AACCTA |
|  | bottom | CACATC | ACGAAG | TGATTG | CTAGGA |
| D* | top | TAGGTT | CCTGTA | TCTGTC | TGTCTC |
|  | bottom | TCCTAG | CAATCA | CTTCGT | GATGTG |
| E | top | GATATG | ATGCAC | TTCCTG | TTCCAT |
|  | bottom | TTCTCA | CTGTGA | TATTCC | AGATTC |
| E* | top | ATGGAA | CAGGAA | GTGCAT | CATATC |
|  | bottom | GAATCT | GGAATA | TCACAG | TGAGAA |
| F | top | GGTAAT | ACTGAG | TCATCC | GGTATT |
|  | bottom | ATTCAG | CTTGAG | CGATTA | ATTCTG |
| F* | top | AATACC | GGATGA | CTCAGT | ATTACC |
|  | bottom | CAGAAT | TAATCG | CTCAAG | CTGAAT |

**Figure S2.** Bond module lookup table. The 4 top and 4 bottom sticky end sequences (5' to 3') are provided for bond module. Any set of 8 sequences can be applied to any core face. As an example, subunits with all sides equipped with the 'sA' sequences (top 6 bps; bottom 5 bps) are good for $T$=1 capsid assembly (shown in Figure S1B orange sequence segment). Consult Figure S3 for how to use the sequences listed in the table to build other assembly structures described in this work. The sA, sB, etc. sequence sets can be used when self-complementary bindings are needed between subunits (with the two jointed subunit faces having the same sequence; Figure S1C *top*). The A-A*, B-B*, etc. sequence sets can be used for binding between two different subunit species (with the two jointed subunit faces having different sequences; Figure S1C *bottom*). Here we provide binding sequences with varying lengths that can be used for desired binding strength. Use black, black + red, black + red + blue, or black + red + blue + purple sequences for sticky ends of 5, 6, 7, or 8 bps long (weak to strong binding), respectively.

| Assembly structure | Subunit | Binding angle (S1, S2, S3) | | Angle module (S1, S2, S3) | Bond module (S1, S2, S3) | |
|---|---|---|---|---|---|---|
| | | | | | Sequence | bp number |
| 2D sheet with contenious tiling | single species | 0°, 0°, 0° | top | 3, 3, 3 poly-T | sA, sA, sA | 5, 5, 5 bps |
| | | | bottom | 3, 3, 3 poly-T | sA, sA, sA | 5, 5, 5 bps |
| 2D sheet with tetramer tiling | yellow species | 0°, 0°, 0° | top | 3, 3, 3 poly-T | A, A*, B | 5, 5, 5 bps |
| | | | bottom | 3, 3, 3 poly-T | A, A*, B | 5, 5, 5 bps |
| | green species | 0°, 0°, 0° | top | 3, 3, 3 poly-T | B*, B*, B* | 5, 5, 5 bps |
| | | | bottom | 3, 3, 3 poly-T | B*, B*, B* | 5, 5, 5 bps |
| $T$=1 capsid shell | single species | 41.8°, 41.8°, 41.8° | top | 14, 14, 14 poly-T | sA, sA, sA | 6, 6, 6 bps |
| | | | bottom | 3, 3, 3 poly-T | sA, sA, sA | 5, 5, 5 bps |
| $T$=4 capsid shell | yellow species | 41.8°, 41.8°, 0° | top | 14, 14, 3 poly-T | A, A*, B | 6, 6, 7 bps |
| | | | bottom | 3, 3, 3 poly-T | A, A*, B | 5, 5, 7 bps |
| | green species | 0°, 0°, 0° | top | 3, 3, 3 poly-T | B*, B*, B* | 7, 7, 7 bps |
| | | | bottom | 3, 3, 3 poly-T | B*, B*, B* | 7, 7, 7 bps |
| Short tube | yellow species | 41.8°, 41.8°, -21.6° | top | 14, 14, 3 poly-T | sA, sB, A | 6, 6, 5 bps |
| | | | bottom | 3, 3, 8 poly-T | sA, sB, A | 5, 5, 5 bps |
| | green species | 41.8°, 41.8°, -21.6° | top | 14, 14, 3 poly-T | B, C, A* | 6, 6, 5 bps |
| | | | bottom | 3, 3, 8 poly-T | B, C, A* | 5, 5, 5 bps |
| | purple species | 41.8°, 41.8°, N/A | top | 14, 14, N/A poly-T | B*, C*, N/A | 6, 6, N/A bps |
| | | | bottom | 3, 3, N/A poly-T | B*, C*, N/A | 5, 5, N/A bps |

**Figure S3.** Angle and bond module lookup table for various assembly structures. Here we provide detailed choices of angle modules (green) and bond modules (orange) to synthesize specific subunits for assembly of 2D sheets, $T$=1 and $T$=4 capsids, and short tubes described in Figure 2. Any chosen angle and bond modules will appear as extension of the conserved 'anchor' segment as detailed in Figure S1B. The sequences of bond modules can be found in Figure S2. See Methods for detailed design strategies.

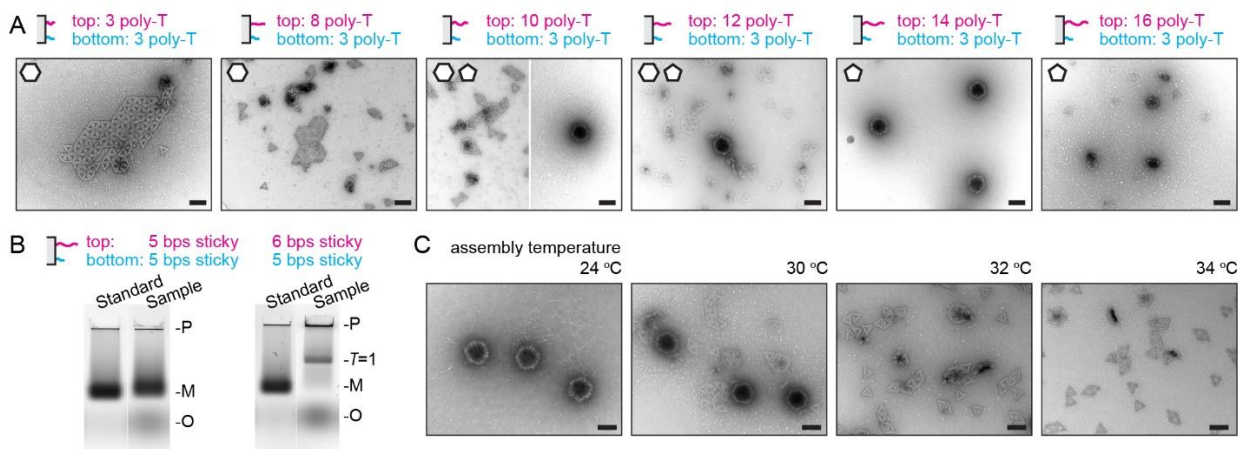

**Figure S4.** Choosing the appropriate angle module, bond module, and temperature for $T$=1 capsid assembly. (A) An appropriate angle module could be determined through systematic trials, with a fixed core module, fixed bond module, fixed $\ell_{bottom}$ = 3 poly-T, and only varying $\ell_{top}$. With gradually-increased $\ell_{top}$, one could observe transition of assembly structures – large 6-fold planar sheet (3 poly-T), small 6-fold planar sheet (8 poly-T), mixture of 6-fold structures and 5-fold $T$=1 capsid shell (10 and 12 poly-T), and highly-specific 5-fold $T$=1 capsid shell (14 poly-T). Note that the $\ell_{top}$ = 16 poly-T case gives highly-specific 5-fold structures (e.g. pentamers), but the binding strength is too weak to form a closed $T$=1 capsid shell; the issue could be resolved by increasing the binding strength of the bond modules, which are not explored here. We thus choose $\ell_{top}$ = 14 poly-T for the optimal $T$=1 capsid assembly. The assembly temperature is 25 °C. Scale bar: 100 nm. (B) An appropriate bond module includes well-designed binding base pair sequences and suitable binding base pair numbers (sticky end lengths). The former was discussed in Figure S2, and the latter could be determined through systematic trials, with a fixed core module, fixed angle module ($\ell_{top}$ = 14 poly-T, $\ell_{bottom}$ = 3 poly-T), fixed bond module sequence, fixed assembly temperature (e.g. 25 °C), and only varying sticky end lengths. The binding strength should be 'just enough' for high-order assembly; below which subunits only form small clusters but no larger structures (*left*, using 5 bps for both top and bottom overhanging ssDNA sticky ends), and above which target assembly can form (*right*, using 6 bps and 5 bps for the top and bottom sticky ends, respectively). The agarose gel is used to characterize product contents, with 'P', '$T$=1', 'M', and 'O' indicate large aggregation in gel pocket, target $T$=1 capsid, monomer, and abandoned oligos, respectively. We thus choose the 6 bps / 5 bps combination for the optimal $T$=1 capsid assembly. (C) Once the subunit structure is determined, the optimal assembly temperature should be picked to be below but near the structure 'melting temperature', above which subunits tend to remain as monomers or small clusters. Here we choose 25 °C. Scale bar: 100 nm.

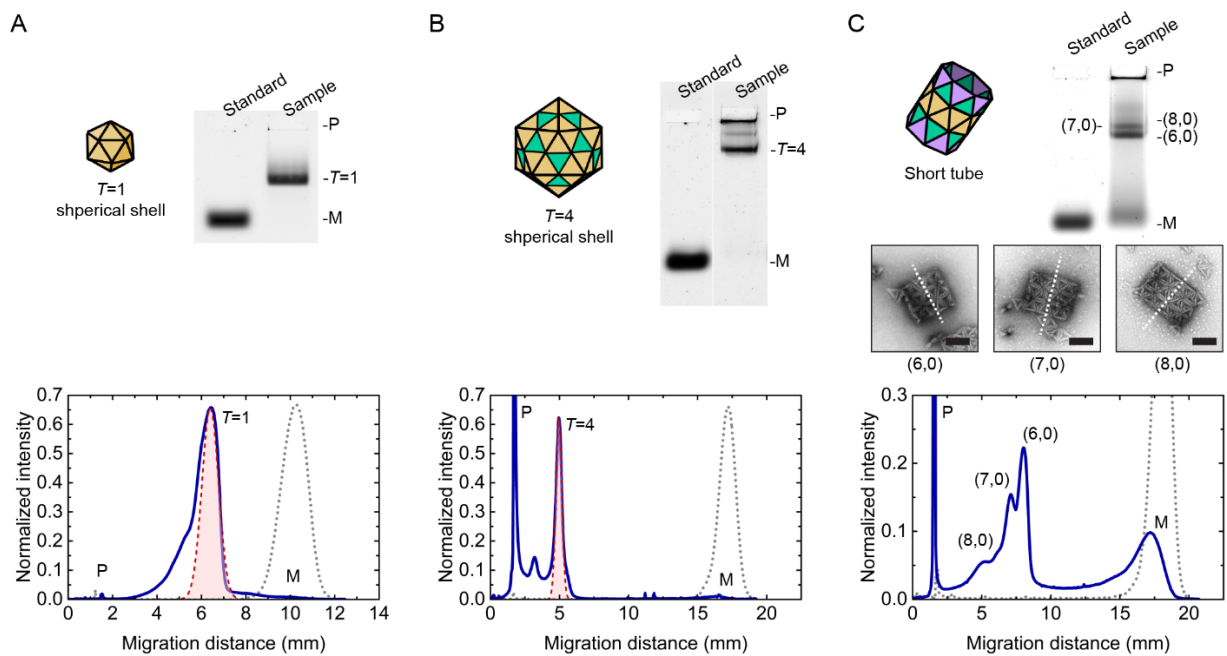

**Figure S5.** Characterizing assembly yield and specificity using gel electrophoresis. (A-C) Assembly products characterized by gel electrophoresis for three exemplary structures, including (A) $T=1$ small spherical shells (see Figure 2C), (B) $T=4$ large spherical shells (see Figure 2D), and (C) short tubes (see Figure 2E). Top row: laser-scanned fluorescent images of agarose gels showing the monomer standard (*left lane*) and the assembly product (*right lane*). The samples are loaded into the top gel pockets and migrate toward the bottom, with smaller species moving faster. The 'P', 'M', '$T=1$', '$T=4$', '(6,0)', '(7,0)', '(8,0)' labels indicate the position of gel pocket (aggregation), peak of monomer population, peak of $T=1$ capsid population, peak of $T=4$ capsid population, and three peaks of short tube populations with varying diameter. The (6,0), (7,0), and (8,0) tubes contain 12,14, and 16 subunits alone its ring direction, respectively, and are shown under EM (the white dashed line indicates the axis of symmetry; scale bar: 100 nm). Bottom row: the intensity profiles obtained from gels show the spatial distribution and relative species population in each sample. Both the assembly (blue solid curves) and the monomer standard (grey dotted curves) are presented. The $T=1$ (A) and $T=4$ (B) capsid population peaks are fitted with a Gaussian (red dashed curves), and the area underneath the Gaussian is defined as the complete capsid yield.

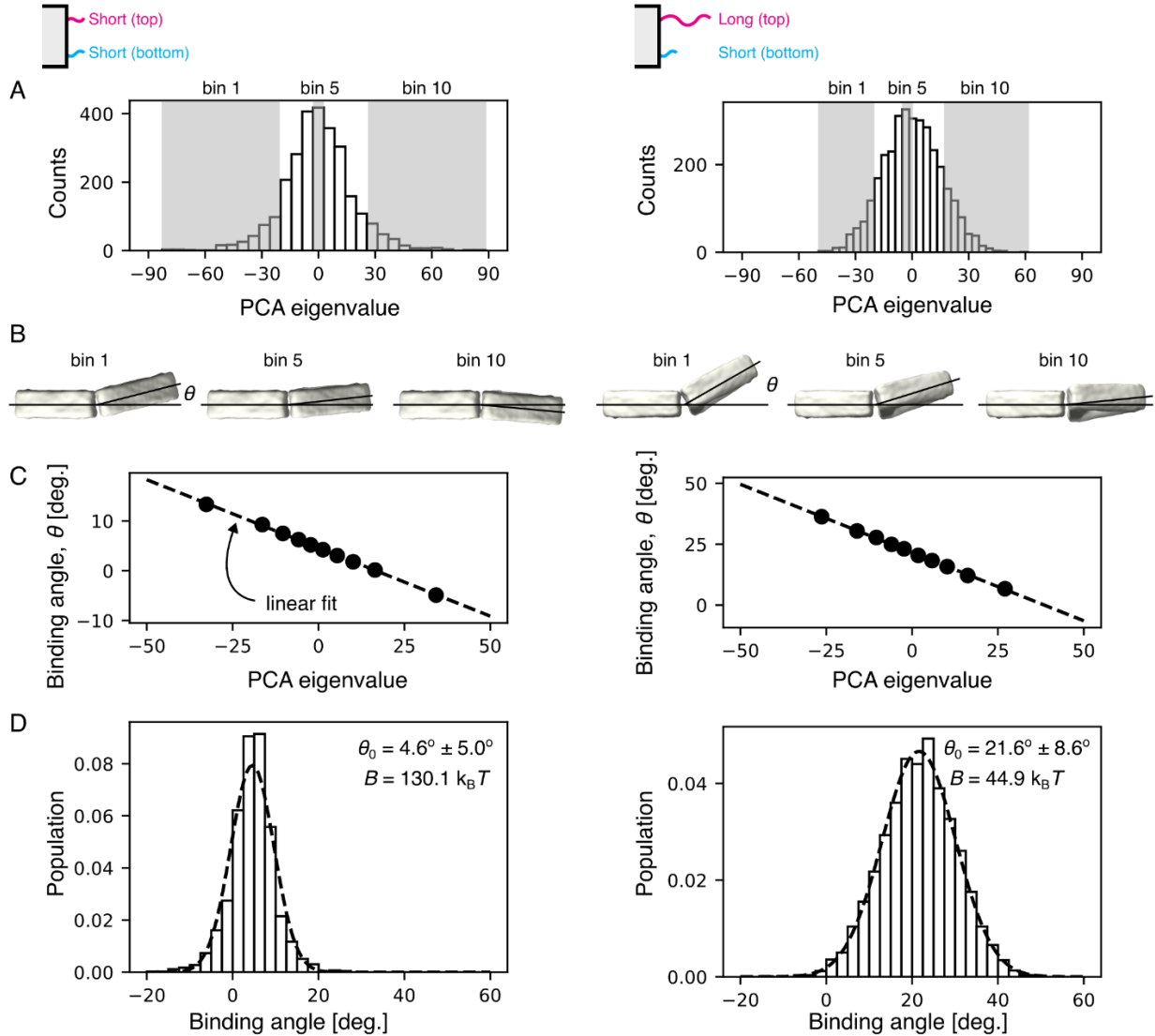

**Figure S6.** Mechanical property of the subunit-subunit joint characterized by cryo-EM. (A) Histogram of eigenvalues corresponding to the multi-body PCA mode that is closest to a bending mode. The distribution in (A) is divided into ten equi-populated bins and the dimer configuration for the average eigenvalues of the PCA mode is output as a density map, shown in (B) for bins 1, 5, and 10. From these dimer configurations we measure the binding angle along the centerline of the two triangles. (C) Plotting the measured angles and the average eigenvalues for each bin shows that they are linearly related. Fitting these points (*dashed line*) allows us to convert the PCA eigenvalues into opening angles for the dimers. (D) Using the linear relationship, we construct distributions for the binding angle of the dimers. We find that these distributions appear normal, so we fit them to a Gaussians (*dashed curve*). This gives us the average binding angle for the dimer. By assuming that these distributions arise from a Boltzmann distribution of the elastic bending energy, we can estimate a bending elastic modulus as $B=1/\sigma^2$, where $\sigma$ is the standard deviation of the angle distribution. *Left column*: data for subunits jointed with angle modules $\ell_{top} = \ell_{bottom} = 3$ poly-T. *Right column*: data for subunits jointed with angle modules $\ell_{top} = 14$ poly-T, $\ell_{bottom} = 3$ poly-T.

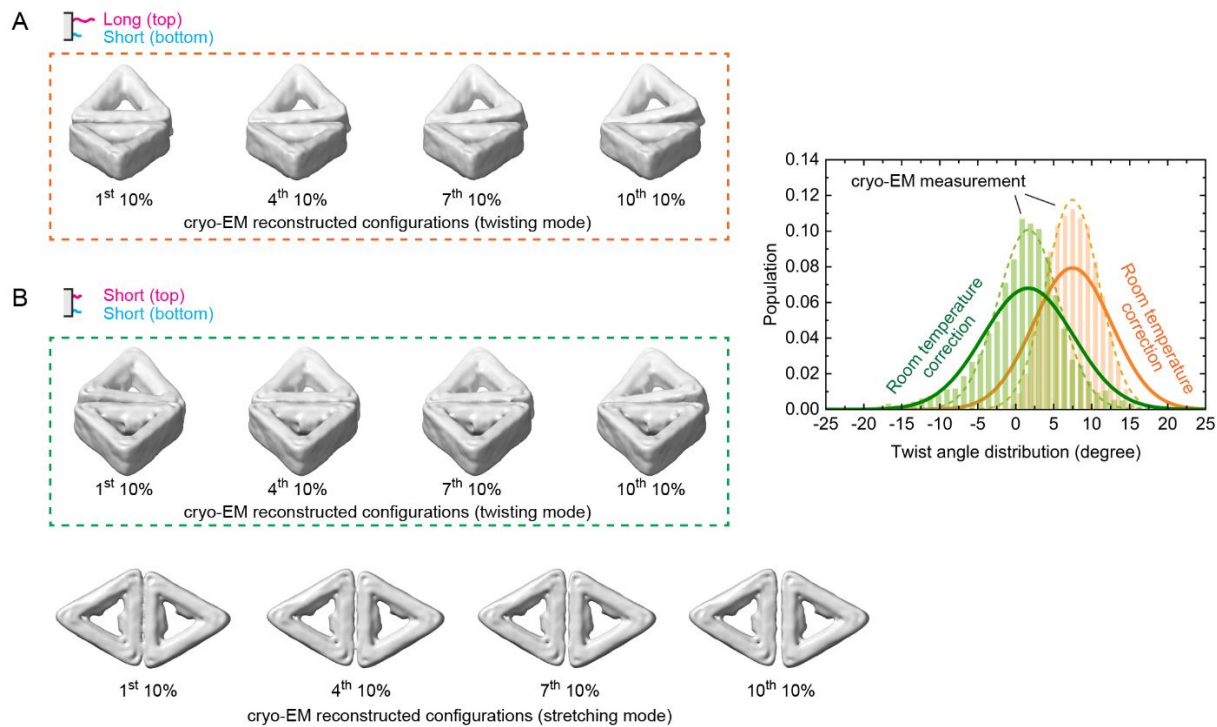

**Figure S7.** Twisting and stretching modes of the subunit-subunit joint characterized by cryo-EM. (A) The dimerized subunit pair with angle modules $\ell_{top} = 14$ poly-T $> \ell_{bottom} = 3$ poly-T twists if freely suspended in solution. The four cryo-EM reconstructions represent average configurations of the most clockwise twisted 10% ensemble (negative twist angle), the most counterclockwise twisted 10% ensemble (positive twist angle), and another two states in between. The twist angle distribution is extracted from the cryo-EM observation (orange bar graph, 136 K), fitted by a Gaussian (dashed orange curve), and rescaled to the room temperature ensemble (solid orange curve, 298 K), giving an estimated twist elastic modulus of 124.1 $k_BT/rad^2$. (B) The dimer joined by angle modules $\ell_{top} = \ell_{bottom} = 3$ poly-T presents both twisting mode with an estimated twist elastic modulus of 92.4 $k_BT/rad^2$ (green curves and top cryo-EM reconstructions) and stretching mode (bottom cryo-EM reconstructions).

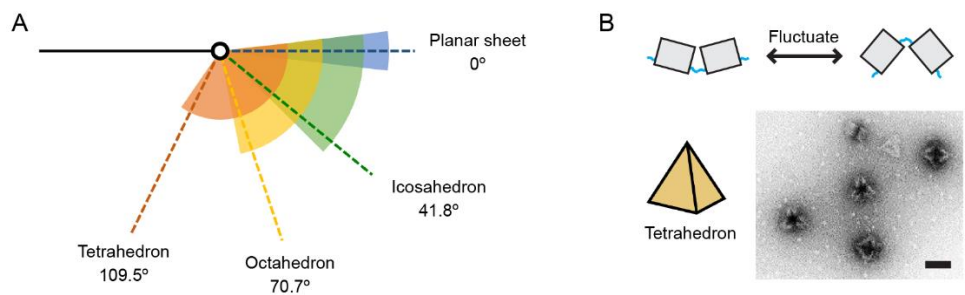

**Figure S8.** Smallest-allowed closed structure is favorable. (A) Schematic illustrating the idea that if there is more than one possible polymorph within the fluctuation range, the closed structure with the smallest number of subunits is kinetically favorable. The blue, green, yellow, and red shaded circular sector indicates the range of binding angle fluctuation that would lead to the corresponding closed structure of planar sheet (case in Figure 2A, Figure 3A), icosahedron (case in Figure 2C, Figure 3B), octahedron, and tetrahedron (case in (B)), respectively. (B) A very fluctuated subunit-subunit joint can be realized by only employing bottom ssDNA strands in the angle modules (*top*). Such subunits self-assemble into tetrahedron as shown in the EM image. Scale bar: 50 nm (*bottom*).

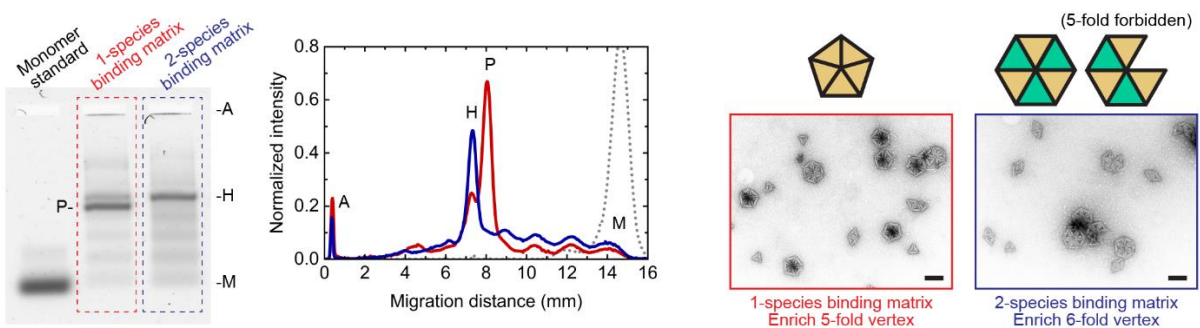

**Figure S9.** Selectively enriching 5-fold or 6-fold structure using subunits with the same flexible angle module. As shown in Figure 4A, the angle module $\ell_{top}$ = 14 poly-T, $\ell_{bottom}$ = 3 poly-T can be programmed to preferentially form different structural binding angles by increasing the number of distinct subunits from one to two. The 1-species binding matrix case (red color-coded) favors forming the 41.8° binding angle that makes a 5-fold vertex. The 2-species binding matrix case (blue color-coded) favors forming the 0° binding angle and 6-fold vertex. The experimental observations are demonstrated in the laser-scanned fluorescent images of agarose gel (*left*; 'A', 'H', 'P', 'M' labels indicate the position of gel pocket/aggregation, peak of hexamer, pentamer, and monomer population, respectively), intensity profiles obtained from the gel (*middle*), and EM images (*right*; scale bar: 100 nm).